\documentclass[sigconf,review=false,natbib=true,anonymous=false]{acmart}
\usepackage{subfigure}
\usepackage{algorithm}
\usepackage[noend]{algpseudocode}
\usepackage{multirow} 
\usepackage{rotating}
\usepackage{tikz}
\usepackage{amsmath}
\usepackage{lipsum,adjustbox}
\usepackage{vcell}
\usepackage{xcolor}
\usepackage{xspace}
\usepackage{diagbox}
\usepackage{pifont}
\usepackage{enumitem}
\usepackage{balance}
\usepackage{cleveref}
\usepackage{placeins}
\usepackage{graphicx}
\usepackage{pgfplots}
\usepackage[normalem]{ulem}

\definecolor{darkgreen}{RGB}{0,128,0} 
\definecolor{darkred}{RGB}{178,34,34}  
\newcommand{\framework}{Lion\xspace}
\copyrightyear{2025}
\acmYear{2025}
\setcopyright{rightsretained}

\makeatother

\usepackage{enumitem}
\newlist{inlinelist}{enumerate*}{1}
\setlist*[inlinelist,1]{%
  label=(\roman*),
}
\usepackage{subfigure}

\newcommand{\ContrastiveAcronym}[0]{CL\xspace}

\title{Scaling Sparse and Dense Retrieval in Decoder-Only LLMs}

\author{Hansi Zeng}
\affiliation{\institution{University of Massachusetts Amherst}
\country{United States}}
\email{hzeng@cs.umass.edu}

\author{Julian Killingback}
\affiliation{\institution{University of Massachusetts Amherst}
\country{United States}}
\email{jkillingback@cs.umass.com}

\author{Hamed Zamani}
\affiliation{\institution{University of Massachusetts Amherst}
\country{United States}}
\email{zamani@cs.umass.edu}

\begin{document}


\begin{abstract}
Scaling large language models (LLMs) has shown great potential for improving retrieval model performance; however, previous studies have mainly focused on dense retrieval trained with contrastive loss (CL), neglecting the scaling behavior of other retrieval paradigms and optimization techniques, such as sparse retrieval and knowledge distillation (KD).
In this work, we conduct a systematic comparative study on how different retrieval paradigms (sparse vs. dense) and fine-tuning objectives (CL vs. KD vs. their combination) affect retrieval performance across different model scales. Using MSMARCO passages as the training dataset, decoder-only LLMs (Llama-3 series: 1B, 3B, 8B), and a fixed compute budget, we evaluate various training configurations on both in-domain (MSMARCO, TREC DL) and out-of-domain (BEIR) benchmarks.
Our key findings reveal that:
(1) Scaling behaviors emerge clearly only with CL, where larger models achieve significant performance gains, whereas KD-trained models show minimal improvement, performing similarly across the 1B, 3B, and 8B scales.
(2) Sparse retrieval models consistently outperforms dense retrieval across both in-domain (MSMARCO, TREC DL) and out-of-domain (BEIR) benchmarks, and they demonstrate greater robustness to imperfect supervised signals. 
(3) We successfully scale sparse retrieval models with the combination of CL and KD losses at 8B scale, achieving state-of-the-art (SOTA) results in all evaluation sets.

\end{abstract}




\maketitle

\section{Introduction}  



Scaling large language models (LLMs) has led to significant improvements across various NLP tasks \cite{gpt-3,chinchilla,mt-scaling-1,mt-scaling-2,emergent-llm}. These scaling benefits also extend to retrieval tasks~\cite{llm2vec,repllama,scale-dr,llama2vec,echo-embed,Tao2024LLMsAA}. Recent studies~\cite{repllama,llama2vec,echo-embed} have demonstrated that dense retrieval models parameterized by large decoder-only LLMs, such as LLaMA2-7B~\cite{llama2}, can significantly outperform traditional BERT-based retrieval models on both in-domain and out-of-domain benchmarks.

These studies mainly focus on dense retrieval, neglecting sparse retrieval~\cite{snrm,splade,splade-v2,spade}—a single-vector retrieval paradigm that uses high-dimension sparse vectors, where each index represents a vocabulary term. An important feature of sparse retrieval models is that each contextualized token embedding is projected into the vocabulary space, where the model can include information about the original token and do term expansion~\cite{doc-exp,query-exp-llm,deepct,splade}. To do this well, the model needs to know information about each token when it does the projection or it will not be able to include all the relevant information. This is a problem for decoder-only LLMs as their causal attention masks are designed for generation where the next token is not known. Thus, the embedding used for the projection of the $i$th token is not aware of the true $i$th token. To address this shortcoming, we adopt the approach from~\cite{llm2vec} by replacing the causal mask with a bidirectional mask~\cite{bert} and doing a masked next-token prediction (MNTP)~\cite{llm2vec} pre-training step. 
This modification minimally alters the decoder-only LLM architecture while making it better suited for sparse retrieval. With this issue resolved, we can now systematically compare sparse and dense retrieval in LLMs to understand their scaling behaviors and effectiveness.


\citet{scale-dr} demonstrated that scaling dense retrieval models reduces contrastive entropy and improves test-set ranking metrics when trained with contrastive loss (CL). However, the scaling behavior of retrieval models trained with knowledge distillation (KD)~\cite{colbertv2,rocketqav2,cl-drd,tas-b,margin-mse,tct-colbert,kd}, another widely used and effective retrieval optimization method, remains underexplored. KD typically employs an external cross-encoder reranker as a teacher model to provide augmented training labels for optimizing student retrieval models~\cite{tas-b}. While KD-trained models have been shown to outperform CL-trained models~\cite{tas-b,cl-drd}, it remains unclear whether this advantage persists as model size increases. As the performance of KD models is inherently tied to the quality of the teacher model, its effectiveness may diminish as models become more capable, making it essential to examine how KD scales with model size.

Furthermore, the scaling behavior of retrieval models in out-of-domain benchmarks was also neglected in previous work~\cite{scale-dr,scale-note,scaling-advertisement-retrieval}. \citet{dragon} has shown that dense retrieval performance in in-domain and out-of-domain settings is often negatively correlated, whereas sparse retrieval exhibits stronger generalization capabilities. This raises important questions: as retrieval models scale, does this trade-off persist or can scaling mitigate the gap, and whether sparse retrieval continues to generalize better than dense retrieval? 

Given these open questions, our work conducts a comprehensive study of how different retrieval paradigms (sparse and dense) and fine-tuning objectives (\ContrastiveAcronym and KD) interact with model scaling.
Specifically, we use the LLaMA3~\cite{llama3} series (1B, 3B, and 8B) as the backbone for our retrieval models. To ensure a fair comparison, we fix the compute budget in all training configurations. For evaluation, we use MS MARCO Dev \cite{msmarco}, TREC DL 2019 \cite{trec-dl-19}, and TREC DL 2020 \cite{trec-dl-20} as in-domain benchmarks, and the BEIR benchmark \cite{beir} for out-of-domain zero-shot evaluation. We employ parameter-efficient fine-tuning via LoRA \cite{lora} for training. Our experiments reveal several key findings:
\begin{itemize}[leftmargin=*]
    \item Unlike contrastive loss, KD-trained models do not exhibit significant performance gains as model size increases. For instance, KD-trained dense retrieval models overfit and underperform their \ContrastiveAcronym-trained counterparts in BEIR at the 3B and 8B scales.
    \item Sparse retrieval models consistently outperform dense retrieval models across all evaluation benchmarks with all the fine-tuning objectives. They demonstrate greater robustness as demonstrated by the out-of-domain BEIR benchmark.
    \item The combination of \ContrastiveAcronym and KD loss achieves the best performance, enabling our 8B sparse retrieval model, named \framework-SP-8B, to become the state-of-the-art model in all the evaluation sets. For instance, \framework-SP-8B outperforms ColBERTv2 by $10.6\%$ in BEIR, and RepLlama by $4.1\%$ in TREC DL 19+20.
\end{itemize}
To improve reproducibility, we open-source our codebase in 
\url{https://github.com/HansiZeng/scaling-retriever}.

\section{Setup}
\begin{table}[t]
    \centering
    \caption{The overview of experimental setup.}
    \vspace{-0.2cm}
    \scalebox{0.85}{
    \begin{tabular}{l!{\color{lightgray}\vrule}l}
    \toprule
        \multicolumn{2}{l}{\textbf{Modeling and Training Objectives: }} \\
        LLM Model Sizes & 1B, 3B, 8B (Llama3 series) \\
        Retrieval Paradigms & dense retrieval, sparse retrieval \\
        Pretraining Objective & MNTP \\
        Finetuning Objectives &  CL, KD, CL+KD \\
        \midrule
        \multicolumn{2}{l}{\textbf{Training and Evaluation Datasets:}} \\       
        Pretrain and Finetuning Corpus & MSMARCO (8.8M docs, 532K queries) \\
        In-domain Evaluation & MSMARCO-Dev, TREC DL 19 \& 20 \\
        Out-of-domain Evaluation & BEIR \\ 
    \bottomrule
    \end{tabular}}
    \label{tab:exp_setup}
    \vspace{-0.5cm}
\end{table}


Our experiments focus on the impact of three variables on in-domain and out-of-domain performance. These variables are: (1) the size of the backbone model. We use the Llama 3 family of models from 1B to 8B (2) the retrieval paradigm including sparse and dense retrieval (3) the fine-tuning objective. The full information on these variables and the training details are presented in Table~\ref{tab:exp_setup}.

\subsection{Retrieval Paradigm}
In this work, we focus on two widely-adopted retrieval paradigms: dense retrieval \cite{ance, cl-drd, dpr} and sparse retrieval \cite{splade, splade-v2,snrm}. These approaches are chosen because they have demonstrated strong retrieval performance \cite{cl-drd, splade-v2,dragon} and both fall under the category of single-vector retrieval, where each document (or query) is encoded into a single vector.

Given the text \( \texttt{T} = [t_1, \ldots, t_L] \) with $L$ tokens and the large language model (LLM) \( M_\theta \) we can get the contextualized sequence representation for the text with hidden dimension \( D \) as follows:
\[
\mathbf{H}_T = M_\theta(\texttt{T}) \in \mathbb{R}^{D \times L}.
\]

For our \textit{dense retrieval} we use mean pooling to obtain a single vector representation:
\[
\vec{\mathbf{h}_T} = \text{AvgPool}(\mathbf{H}_T) \in \mathbb{R}^D.
\]

For \textit{sparse retrieval}, we project \( \mathbf{H}_T \) into the high-dimensional subword token space by doing a matrix multiplication with the LLM's embedding table \( \mathbf{E} \in \mathbb{R}^{D \times V} \) where \( V \) is the vocabulary size, followed by a max-pooling and rescaling operation similar to \cite{splade-v2}:
\[
\vec{\mathbf{c}_T} = \log \big( 1 + \text{ReLU} \big( \text{MaxPool} (\mathbf{E}^T \cdot \mathbf{H}_T) \big) \big) \in \mathbb{R}^V.
\]
To sparsify the high-dimensional vector \( \vec{\mathbf{c}_T} \), we apply FLOP regularization \cite{flop-reg} during training. 

For dense and sparse retrieval the relevance score, \( s( \texttt{q},  \texttt{d}) \), between a query \( \texttt{q} \) and a document \( \texttt{d} \) is computed in a unified way. Given the corresponding vector representations \( \vec{\mathbf{q}} \) and \( \vec{\mathbf{d}} \) the relevance score is defined as:
\begin{align} \label{eq:rel_score}
    s( \texttt{q},  \texttt{d}) = \vec{\mathbf{q}} \cdot \vec{\mathbf{d}}
\end{align}

These two paradigms represent textual information differently: sparse retrieval captures \textit{lexical matching signals}, while dense retrieval focuses on \textit{semantic matching signals}. This difference makes them an ideal pair for comparative studies.

\subsection{Pretraining}
To adapt the decoder-only LLMs for the retrieval tasks, we adopt the pretraining strategy from~\cite{llm2vec}, which consists of two key components: (1) enabling bidirectional attention and (2) applying masked next token prediction (MNTP).

\paragraph{Enabling Bidirectional Attention} 


We first replace the causal attention mask with a bidirectional attention mask. This allows each token to obtain information from every other token and itself, thus improving the ability to represent the text content and increasing the retrieval performance~\cite{echo-embed,llm2vec}. 

\paragraph{Masked Next Token Prediction (MNTP)}


To adapt the LLMs to bidirectional attention, we train the models with masked next token prediction (MNTP) \cite{llm2vec}. Specifically, given an input sequence $x = (x_1, x_2, \dots, x_N)$, we mask $20\%$ of the input tokens and then train the model to predict these masked tokens based on the past and future context. Crucially, when predicting a masked token at position $i$, we calculate the loss based on the logits obtained from the token representation at the previous position $i - 1$, instead of from the masked position itself. This is done because it aligns with the LLM's existing causal training. We use MS MARCO corpus as pre-training corpus.\footnote{We do at most 10,000 pre-training steps on MS MARCO, which can be completed in 17 hours on 2 A100 GPUs for Llama3-8B.}

\subsection{Finetuning}
We choose (1) Contrastive Loss (\ContrastiveAcronym), (2) Knowledge Distillation (KD), and (3) the combination of \ContrastiveAcronym and KD as our retrieval fine-tuning objectives.

\ContrastiveAcronym is a simple yet effective retrieval training objective. It does not rely on any external models during training, and has been proven effective on both in-domain and out-of-domain evaluation sets \cite{co-condenser}. Formally, given a query $q$, a positive document $d^+$, and a list of negative documents $\mathcal{D}_q^- := [ d^-_1, \ldots , d^-_N]$, 
using Eq \eqref{eq:rel_score}, the \ContrastiveAcronym loss for the query $q$ is:
\[
\mathcal{L}_{\text{\ContrastiveAcronym}} = -\log \frac{\exp(s(q,d^+))}{\exp(s(q,d^+) + \displaystyle\sum_{d^- \in \mathcal{D}_q^-} \exp(s(q, d^-))}
\]


\textit{KD} leverages an external teacher model\footnote{We use \url{https://huggingface.co/cross-encoder/ms-marco-MiniLM-L-6-v2}.} - typically a cross-encoder reranker - to provide more fine-grained soft labels for retrieval model training. 
We use MarginMSE \cite{margin-mse} as the KD loss in our initial comparison due to its strong performance. 
Formally, given a query $q$, a positive document $d^+$, a negative document $d^-$, and teacher scores $T(q, d^+)$ and $T(q, d^-)$, the MarginMSE loss is:
\[ 
    \mathcal{L}_{KD} = MSE(s(q, d^+) - s(q,d^-),\, T(q,d^+) - T(q,d^-)\big)
\]


We also explore combining \textit{\ContrastiveAcronym} and \textit{KD}. As \ContrastiveAcronym works with  a list of documents and MarginMSE only works for triplets we switch the KD loss to KL-Divergence \cite{colbertv2,rocketqav2}. KL-Divergence minimizes the difference in the listwise distribution of the teacher and student. The final loss is the weighted sum of the \ContrastiveAcronym and KD loss terms which is: $\mathcal{L}_{\text{comb}} = 0.5 \cdot (\mathcal{L}_{\text{CL}} + \mathcal{L}_{KD} )$.

\subsection{Training and Evaluation}
We use the MS MARCO passage retrieval dataset~\cite{msmarco} for training. For in-domain evaluation, we select MS MARCO Dev \cite{msmarco}, TREC Deep Learning (DL) Track 19 and 20~\cite{trec-dl-19,trec-dl-20}, as they share the same distribution as the training data. We follow the official evaluation protocol, reporting MRR@10 for MS MARCO Dev and nDCG@10 for TREC DL 19 and 20. For out-of-domain evaluation, we adopt the BEIR benchmark \cite{beir}. Following previous work  \cite{colbertv2,dragon,repllama}, we compute nDCG@10 across 13 datasets in BEIR, making the results directly comparable.

\section{Analysis}
\begin{figure}
    \centering
    \includegraphics[width=1\linewidth]{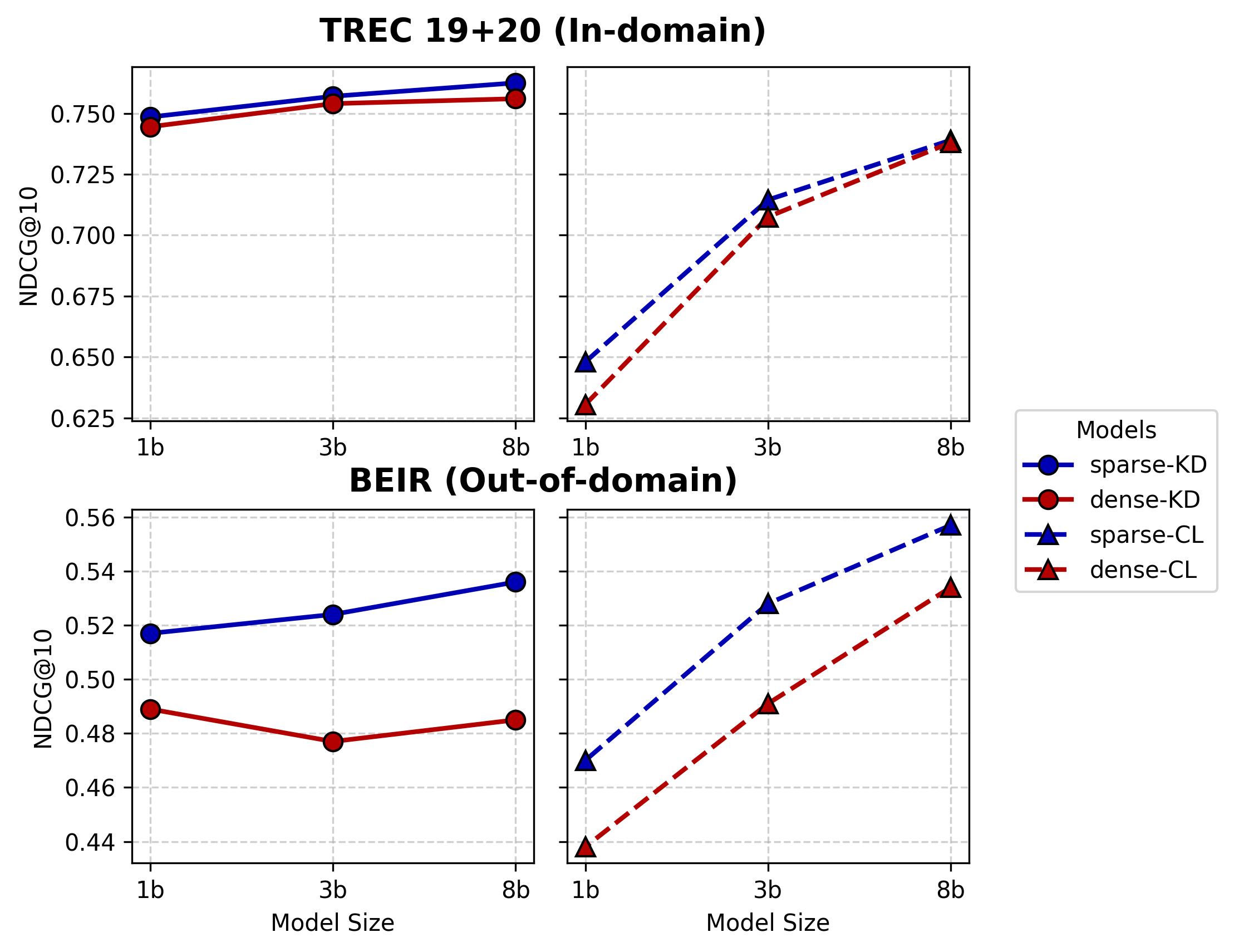}
    \caption{Dense and sparse retrieval results on the combined of TREC DL 19 and 20 (in-domain), and BEIR (out-of-domain) datasets. 
    }
    \label{fig:sparse-vs-dense}
    \vspace{-0.5cm}
\end{figure}

\paragraph{\textbf{Scaling Behavior Clearly Emerges with \ContrastiveAcronym, Not with KD}}
From Figure~\ref{fig:sparse-vs-dense}, we observe that when the fine-tuning objective is \ContrastiveAcronym the retrieval model's performance, both for sparse and dense retrieval, increases significantly as the model size grows for both in-domain and out-of-domain evaluation sets.

In contrast, when KD loss is used the performance improvement is far less pronounced with a decrease in performance in some cases. On the in-domain evaluation set, the 1B and 8B KD models perform similarly. Though, compared to \ContrastiveAcronym the KD models perform much better on the in-domain evaluation sets. For example, the 1B KD models outperform the 8B \ContrastiveAcronym models on TREC 19 and 20. However, when evaluated on the out-of-domain BEIR benchmark, KD-models exhibit signs of overfitting. Specifically, dense-KD models show a decline in performance at larger scales. With both sparse and dense KD models performing worse than \ContrastiveAcronym models on out-of-domain evaluation metrics at larger scales. For example, at the 8B scale, both sparse-KD and dense-KD perform worse than their respective sparse-\ContrastiveAcronym and dense-\ContrastiveAcronym models.
\paragraph{\textbf{Sparse Retrieval consistently Outperforms Dense Retrieval}}

From Figure~\ref{fig:sparse-vs-dense}, we observe that sparse retrieval consistently outperforms dense retrieval with both KD and \ContrastiveAcronym objectives, with a more pronounced increase on the out-of-domain evaluation. For example, at the 8B scale, sparse-KD surpasses dense-KD by $10.5\%$, while sparse-\ContrastiveAcronym outperforms dense-\ContrastiveAcronym by $4.3\%$ in BEIR.

Moreover, unlike dense-KD, which suffers from overfitting to the teacher model, sparse-KD improves steadily as model size increases. While both paradigms perform similarly on in-domain benchmarks, sparse retrieval demonstrates significantly stronger generalization, achieving much better results in zero-shot settings.

\subsection{\ContrastiveAcronym + KD achieves the best Trade-off}
\begin{table}[t]
    \centering
    \caption{Sparse and dense performance with \ContrastiveAcronym + KD loss. Each decimal is the main metric for the associated evaluation dataset. The first number in parentheses indicates the relative change from only KD while the second is the relative change from only \ContrastiveAcronym.} 
    \scalebox{0.85}{\begin{tabular}{l!{\color{lightgray}\vrule}ccc}
    \toprule
       Model Size  & 1B & 3B & 8B \\
        \midrule
        \multicolumn{2}{l}{\textbf{Sparse Retrieval with \ContrastiveAcronym + KD}} \\
        MSMARCO Dev 
        & .410 (\textcolor{darkgreen}{3.8\%}, \textcolor{darkgreen}{22.0\%}) 
        & .417 (\textcolor{darkgreen}{5.0\%}, \textcolor{darkgreen}{7.5\%}) 
        & .417 (\textcolor{darkgreen}{4.2\%}, \textcolor{darkgreen}{0.5\%})
        \\       
        TREC 19+20 & .749 (\textcolor{darkgreen}{0.1\%}, \textcolor{darkgreen}{18.9\%}) & 
.759 (\textcolor{darkgreen}{0.2\%}, \textcolor{darkgreen}{6.2\%}) & 
.762 (\textcolor{darkred}{-0.1\%}, \textcolor{darkgreen}{3.2\%})
 \\ 
        BEIR & .535 (\textcolor{darkgreen}{3.5\%}, \textcolor{darkgreen}{22.1\%}) & 
        .544 (\textcolor{darkgreen}{3.8\%}, \textcolor{darkgreen}{3.0\%}) & 
        .552 (\textcolor{darkgreen}{3.0\%}, \textcolor{darkred}{-0.9\%}) \\ 
        \midrule
        \multicolumn{2}{l}{\textbf{Dense Retrieval  with \ContrastiveAcronym + KD}} \\
        MSMARCO Dev & .404 (\textcolor{darkgreen}{4.1\%}, \textcolor{darkgreen}{20.2\%}) & 
.414 (\textcolor{darkgreen}{5.6\%}, \textcolor{darkgreen}{5.9\%}) & 
.417 (\textcolor{darkgreen}{5.0\%}, \textcolor{darkred}{-0.7\%})
 \\ 
        TREC 19+20 & .749 (\textcolor{darkgreen}{0.6\%}, \textcolor{darkgreen}{18.8\%}) & 
.760 (\textcolor{darkgreen}{0.8\%}, \textcolor{darkgreen}{7.4\%}) & 
.757 (\textcolor{darkgreen}{0.2\%}, \textcolor{darkgreen}{2.7\%}) \\
        BEIR & .500 (\textcolor{darkgreen}{2.2\%}, \textcolor{darkgreen}{14.2\%}) & 
.496 (\textcolor{darkgreen}{4.0\%}, \textcolor{darkgreen}{1.0\%}) & 
.501 (\textcolor{darkgreen}{3.3\%}, \textcolor{darkred}{-6.2\%})
 \\
 \bottomrule
    \end{tabular}}
    \label{tab:nce+kd result}
\end{table}
From Figure~\ref{fig:sparse-vs-dense}, we observe that KD loss significantly boosts the performance of small-scale (1B) retrieval models, while \ContrastiveAcronym provides greater benefits for large-scale (8B) models. 
We hypothesize that small models benefit from the augmented soft labels provided by the teacher model. However, these soft labels are not perfect, and when the model size reaches 8B, the retrieval model's capacity may surpass that of the cross-encoder teacher model. At this point, \ContrastiveAcronym loss becomes more effective.

Building on these findings, we investigate a combination of \ContrastiveAcronym and KD, named \ContrastiveAcronym + KD, that combines the strengths of both, with results shown in Table~\ref{tab:nce+kd result}. We find that \ContrastiveAcronym + KD achieves the best performance balance. For all 1B and 3B models and all evaluation datasets \ContrastiveAcronym + KD increases performance over only KD and only \ContrastiveAcronym. For 8B parameter models there are some decreases in performance but these are offset by gains in other domains in most cases. For instance at the 8B scale, sparse retrieval with the combined loss outperforms the \ContrastiveAcronym-only model by $0.5\%$ on MSMARCO Dev and $3.2\%$ on TREC 19+20, while incurring only a $0.9\%$ loss on BEIR.

\subsection{Sparse Retrieval might Beat the Teacher}
Figure~\ref{fig:sparse-vs-dense} shows that sparse-\ContrastiveAcronym and dense-\ContrastiveAcronym outperform their KD counterparts on BEIR at the 3B and 8B scales. We hypothesize that large retrieval models may have greater capacity than their cross-encoder teacher models. To explore this possibility, we used the teacher model to rerank the original rankings of each retrieval model on BEIR and computed the relative performance difference between the original and reranked results, presented in Table~\ref{tab:sparse_dense_wrt_reranker}.

We observe that sparse-\ContrastiveAcronym surpasses the teacher at both the 3B and 8B scales, while dense-\ContrastiveAcronym outperforms the teacher only at the 8B scale. Another finding is that when KD loss is incorporated (KD and \ContrastiveAcronym + KD), sparse retrieval remains highly robust and consistently outperforms the teacher model. However, for dense retrieval trained with the mixture of losses, performance consistently falls behind the teacher reranker by $1.8\%$, $2.4\%$, and $1.4\%$ at the 1B, 3B, and 8B scales, respectively. These findings suggest that sparse retrieval is a more effective and robust retrieval paradigm and has the potential to serve as a stronger teacher model.
\begin{table}
    \centering
    \caption{nDCG@10 of sparse and dense retrieval finetuned with \ContrastiveAcronym, KD or \ContrastiveAcronym+KD objective on the BEIR benchmark. The relative change from the teacher reranked results is shown in the parentheses.}
    \begin{tabular}{l!{\color{lightgray}\vrule}ccc}
    \toprule
       Model Size  &  1B & 3B & 8B \\
       \midrule
       \multicolumn{3}{l}{\textbf{Sparse Retrieval in BEIR}} \\
        \ContrastiveAcronym & .496 (\textcolor{darkred}{-8.8\%}) & .528 (\textcolor{darkgreen}{1.3\%}) & .557 (\textcolor{darkgreen}{7.4\%})
 \\
 KD & .517 (\textcolor{darkgreen}{0.9\%}) & .524 (\textcolor{darkgreen}{1.9\%}) & .536 (\textcolor{darkgreen}{4.0\%}) \\
    \ContrastiveAcronym + KD & .535 (\textcolor{darkgreen}{4.3\%}) &  .544 (\textcolor{darkgreen}{6.1\%}) & .552 (\textcolor{darkgreen}{7.6\%}) \\
\midrule
\multicolumn{2}{l}{\textbf{Dense Retrieval in BEIR }}  \\
\ContrastiveAcronym & .435 (\textcolor{darkred}{-14.7\%}) &  .491 (\textcolor{darkred}{-5.1\%}) &  .534 (\textcolor{darkgreen}{3.1\%}) \\
KD & .489 (\textcolor{darkred}{-4.5\%}) & .477 (\textcolor{darkred}{-6.7\%}) & .485 (\textcolor{darkred}{-5.3\%})\\
\ContrastiveAcronym + KD & .500 (\textcolor{darkred}{-1.8\%}) &  .496 (\textcolor{darkred}{-2.4\%}) &  .501 (\textcolor{darkred}{-1.4\%}) \\
\bottomrule
    \end{tabular}
    \label{tab:sparse_dense_wrt_reranker}
    \vspace{-.5cm}
\end{table}

\begin{table}[]
    \centering
    \caption{Results of the proposed \framework with other SOTA retrieval models. We use paired t-test with Bonferroni correction with p$\_$value < 0.01. The superscripts refer to significant improvements over CL-DRD($*$), SPLADE++($\dag$), ColBERTv2 ($\ddag$), and RepLlama ($\S$).}
    \scalebox{0.85}{
    \begin{tabular}{l!{\color{lightgray}\vrule}l!{\color{lightgray}\vrule}l!{\color{lightgray}\vrule}l!{\color{black}\vrule}l}
    \toprule
     & \multicolumn{3}{c!{\color{black}\vrule}}{In Domain} & O.O.D \\
     \midrule
    & \textbf{MARCO Dev} & \multicolumn{1}{l}{\textbf{TREC-19}} & \textbf{TREC-20} & \textbf{BEIR-13} \\
    & \multicolumn{1}{c!{\color{lightgray}\vrule}}{ MRR@10} & \multicolumn{2}{c!{\color{black}\vrule}}{ nDCG@10} &  nDCG@10 \\
    \midrule 
    \multicolumn{3}{l}{\textbf{Model size $\leq$ 1B }} \\
    CL-DRD & .381 & .725 & .683 & .448 \\
    SPLADE++ & .389 & .743 & .718 & .503 \\
    ColBERTv2 & .397 & .750 & .746 & .499 \\
    \framework-DS-1B & .404$^{* \dag \ddag}$ & .757$^{* \dag \ddag}$ & .740$^{* \dag}$ & .500$^{* \dag}$ \\
    \framework-SP-1B & .410$^{* \dag \ddag}$ & .747$^{* \dag}$ & .751$^{* \dag}$ & .535$^{* \dag \ddag}$ \\
    \midrule
    \multicolumn{3}{l}{\textbf{Model size > 1B}} \\
    RepLlama & .412 & .743 & .721 & .551 \\
    \framework-DS-8B & \textbf{.417}$^{* \dag \ddag \S}$ & .755$^{* \dag \ddag \S}$ & .759$^{* \dag \ddag \S}$ & .501$^{* \dag \ddag}$ \\
    \framework-SP-8B & \textbf{.417}$^{* \dag \ddag \S}$ & \textbf{.758}$^{* \dag \ddag \S}$ & \textbf{.766}$^{* \dag \ddag \S}$ & \textbf{.552}$^{* \dag \ddag}$ \\
    \bottomrule
    \end{tabular}}
    \label{tab:result-comparison}
    \vspace{-1.cm}
\end{table}

\section{Performance Comparison with Other Retrieval Models}
For comparison with our 1B models we select state-of-the-art baselines from each retrieval paradigm that are smaller than 1B parameters: CL-DRD~\cite{cl-drd} (dense retrieval), SPLADE++~\cite{splade-v2} (sparse retrieval), and ColBERTv2~\cite{colbertv2} (multi-vector retrieval). For models larger than 1B, we use RepLLaMA~\cite{repllama} as the baseline. We refer to our models trained with KD + CL as \framework-[TYPE]-[SIZE] where the type SP indicates a sparse model and DS indicates a dense model, e.g. \framework-SP-8B is the sparse model with LLaMA3-8B as the base model. The in-domain and out-of-domain performance of all models is shown in Table~\ref{tab:result-comparison}. 

We observe that \framework-SP-1B and \framework-DS-1B achieve the best overall results among baselines with $\leq$ 1B parameters, demonstrating that recent large decoder-only LLMs serve as a powerful backbone for document retrieval. For instance, \framework-SP-1B surpasses ColBERTv2 by $3.2\%$ on MS MARCO Dev and by $7.2\%$ on BEIR. Moreover, \framework-SP-8B outperforms the state-of-the-art model RepLLaMA across both in-domain and out-of-domain benchmarks, underscoring the effectiveness of the sparse retrieval paradigm and the combination of CL and KD loss. 

Additionally, \framework's performance on individual BEIR datasets is provided in Table~\ref{tab:beir-per-ds-results} in the Appendix for reference.
\vspace{-.25cm}
\section{Related Work}
 Single-vector retrieval models can be broadly categorized into two types: dense retrieval \cite{ance,dpr,cl-drd,rocketqav2,margin-mse,tas-b,repllama,llama2vec,echo-embed} and sparse retrieval \cite{splade,splade-mistral,splade-v2,snrm,spade}. The key distinction lies in their representations: dense retrieval encodes texts into low-dimensional vectors (i.e. 768 dim in BERT) and employs approximate nearest neighbor (ANN)~\cite{ance,faiss-gpu} search for efficient retrieval. In contrast, sparse retrieval represents texts using high-dimensional sparse vectors and relies on an inverted index~\cite{spade,inverted-index} for fast retrieval. 
 
The most straightforward fine-tuning objective for retrieval models is supervised contrastive loss (CL) \cite{repllama,co-condenser}. By incorporating techniques such as hard negative sampling \cite{ance,adore} and retrieval-oriented continued pre-training \cite{b-prop,co-condenser}, the performance of retrieval models can be further improved. 
Knowledge distillation (KD)~\cite{colbertv2,rocketqav2,cl-drd,tas-b,margin-mse,tct-colbert,kd} is another widely used fine-tuning objective. 
It has also been shown to perform comparably to or even better than CL for small-scale retrieval models, particularly on in-domain benchmarks \cite{tas-b,cl-drd}.

Scaling LLMs has consistently led to improved performance across various NLP tasks~\cite{gpt-3,chinchilla,mt-scaling-1,mt-scaling-2,emergent-llm}. This trend also extends to retrieval models. ~\citet{scale-dr} demonstrated that scaling BERT-based dense retrieval models enhances retrieval performance.
Recent works~\cite{llama2vec,echo-embed,splade-mistral,llm2vec} have further explored this trend by training dense retrieval models with larger backbones, such as LLaMA2-7B~\cite{llama2}. These models have achieved state-of-the-art performance, surpassing traditional BERT-based retrieval models.
\vspace{-.25cm}
\section{Conclusions and future work}
We conduct a comprehensive study on the scaling behaviors of different retrieval paradigms and fine-tuning objectives in decoder-only LLMs. One limitation of this work is that we only evaluate knowledge distillation (KD) with a single teacher model. In future work, we aim to expand this analysis by exploring teacher models with varying sizes and different supervision strategies to gain deeper insights into their impact on retrieval performance.

\begin{acks}
    This work was supported in part by the Center for Intelligent Information Retrieval, and in part by the Office of Naval Research contract number N000142412612. Any opinions, findings and conclusions or recommendations expressed in this material are those of the authors and do not necessarily reflect those of the sponsor.
\end{acks}

\bibliographystyle{ACM-Reference-Format}
\bibliography{XX-references}
\clearpage

\appendix
\section{Implementation Details}
\begin{table}[t]
    \centering
    \caption{Training configurations for sparse and dense retrieval models. 
    "Grad. accu." refers to gradient accumulation steps, and "Num. negs" represents the number of negative documents used during training.}

    \scalebox{0.85}{
    \begin{tabular}{l!{\color{lightgray}\vrule}ccccc}
    \toprule
     Loss type & Size & Epochs & Batch size & Grad. accu. & Num. negs \\
        \midrule
        CL, CL+KD & 1B & 7 & 28 & 1 & 16 \\
        CL, CL+KD & 3B & 3 & 16 & 2 & 16 \\
        CL, CL+KD & 8B & 1 & 8 & 4 & 16 \\
        \midrule
        KD & 1B & 56 & 256 & 1 & 1 \\
        KD & 3B & 24 & 128 & 1 & 1 \\
        KD & 8B & 8 & 64 & 2 & 1 \\
        \toprule
    \end{tabular}}
    \label{tab:train-config}
\end{table}
For MNTP pre-training, we use 2×A100 80GB GPUs. The training batch size per device is 32, with a maximum sequence length of 512 and a gradient accumulation step of 1. We apply BERT’s masking strategy with a masking probability of $20\%$ and train all models for 10,000 steps.

For fine-tuning, we use 4×A100 80GB GPUs for all training runs and maintain a fixed compute budget across all configurations. 
Each training run completes within approximately 40–44 hours. 
The key fine-tuning hyperparameters that impact the training time for different model scales and fine-tuning objectives are presented in Table~\ref{tab:train-config}.
As for other hyperparameters, we set the learning rate at 1e-4, the warmup ratio at $4\%$ of all training steps, the maximum query length at 64, the maximum document length at 128. We set LoRA $r=16$ and $\alpha=32$. In sparse retrieval training, the FLOPs regularization coefficient is 0.05 for query vectors and 0.04 for document vectors. We use AdamW as the optimizer. For BEIR evaluation, the maximum sequence length is set to 512.

\section{BEIR results}
\begin{table}[t]
    \centering
    \caption{retrieval performance across datasets in BEIR.}
    \scalebox{0.85}{\begin{tabular}{l!{\color{lightgray}\vrule}cccc}
        \toprule
        Dataset & \framework-DS-1B & \framework-SP-1B & \framework-DS-8B & \framework-SP-8B \\
        \midrule
        ArguAna & 0.476 & 0.488 & 0.494 & 0.484 \\
        FiQA & 0.365 & 0.378 & 0.386 & 0.398 \\
        NFCorpus & 0.341 & 0.365 & 0.363 & 0.370 \\
        Quora & 0.766 & 0.791 & 0.726 & 0.874 \\
        SciDocs & 0.177 & 0.171 & 0.191 & 0.178 \\
        SciFact & 0.691 & 0.732 & 0.687 & 0.730 \\
        TREC-COVID & 0.797 & 0.783 & 0.807 & 0.825 \\
        Webis-Touche2020 & 0.274 & 0.330 & 0.232 & 0.358 \\
        Climate-Fever & 0.274 & 0.304 & 0.283 & 0.281 \\
        DBPedia-Entity & 0.442 & 0.464 & 0.456 & 0.465 \\
        FEVER & 0.691 & 0.870 & 0.627 & 0.874 \\
        HotpotQA & 0.625 & 0.692 & 0.658 & 0.729 \\
        NQ & 0.575 & 0.585 & 0.606 & 0.611 \\
        \midrule
        Avg & 0.500 & 0.535 & 0.501 & 0.552 \\
        \bottomrule
    \end{tabular}}
    \label{tab:beir-per-ds-results}
\end{table}

We also report the detailed BEIR performance per dataset for our proposed \framework models shown in Table~\ref{tab:beir-per-ds-results}.

\end{document}